\documentclass[12pt,nofootinbib,preprint,
noshowkeys,noshowpacs,longbibliography,superscriptaddress,tightenlines,floatfix]{revtex4-2}
\usepackage{amsmath,amsfonts,amsthm,amssymb}
\usepackage[dvips]{graphics,graphicx}
\usepackage[usenames,dvipsnames]{color}
\definecolor{darkblue}{RGB}{0,0,196}
\definecolor{darkgreen}{RGB}{0,120,0}
\usepackage[colorlinks=true,linktocpage=true,linkcolor=darkblue,citecolor=red,urlcolor=darkblue]{hyperref}
\usepackage{cancel}
\usepackage{multirow}
\usepackage{longtable}
\usepackage{color}
\usepackage[normalem]{ulem}
\usepackage{hyperref}
\usepackage{bigints}
\usepackage{xparse}
\usepackage{physics}
\usepackage{verbatim}
\usepackage{minibox}
\usepackage{comment}
\usepackage{appendix}
\usepackage{marginnote}
\usepackage{graphicx}
\usepackage[nice]{nicefrac}
\usepackage{pifont}
\def\HP{\hphantom{\alpha}} 

\usepackage{cleveref}
\crefname{equation}{}{eqs.}

\newcommand{\bel}[1]{\begin{eqnarray}\label{#1}}
	\newcommand{\eel}{\end{eqnarray}}
\def\beq{\begin{eqnarray}}
	\def\eeq{\end{eqnarray}}
\newcommand{\p}{\partial}
\newcommand{\rfn}[1]{(\ref{#1})}

\def\HP{\hphantom{\alpha}} 

\newcommand{\sh}[1]{\sinh#1}
\newcommand{\ch}[1]{\cosh#1}

\newcommand{\lab}[1]{\label{#1}}
\def\nn{\nonumber}
\newcommand{\refb}[1]{(\ref{#1})}
\def\cA{{B}}
\def\cB{{C}}
\def\cC{{A}}

\def\cN{{n}}
\def\cE{{\varepsilon}}
\def\cP{{P}}

\def\be{\begin{equation}}
\def\ee{\end{equation}}
\def\ba{\begin{eqnarray}}
\def\ea{\end{eqnarray}}   

\def\a{\alpha}
\def\b{\beta}
\def\g{\gamma}
\def\d{\delta}

\def\half{\frac{1}{2}}

\begin{document}

\preprint{}

\title{Spin hydrodynamics on a hyperbolic expanding background}

\author{Rajeev Singh}
\email{rajeev.singh@e-uvt.ro}
\affiliation{Department of Physics, West University of Timisoara, Bulevardul~Vasile P\^arvan 4, Timisoara 300223, Romania}
\author{Alexander Soloviev}
\email{alexander.soloviev@fmf.uni-lj.si}
\affiliation{Faculty of Mathematics and Physics, University of Ljubljana, Jadranska ulica 19, SI-1000, Ljubljana, Slovenia}
\date{\today} 
\bigskip
\bigskip
\begin{abstract}
We study relativistic spin hydrodynamics on the hyperbolic $\kappa=-1$ flow background recently identified by Grozdanov. This background corresponds to an $SO(2,1)$-invariant, transversely expanding solution with finite spacetime support in Minkowski space, in contrast to the well-known Gubser flow $(\kappa=+1)$ which possesses $SO(3)$ symmetry and infinite transverse extent. Working within the formulation of perfect-fluid spin hydrodynamics, we derive the exact evolution equations for all spin components of the spin potential on the $\kappa=-1$ background. We find that the enhanced early-time expansion rate and the presence of a causal edge lead to a stronger localization of spin dynamics compared to the Gubser case. Remarkably, the azimuthal component of the spin potential oscillates as it decays in the forward lightcone, in stark contrast to the Gubser flow. Thus, our results establish the $\kappa=-1$ flow as a distinct and physically meaningful benchmark for studying spin dynamics in expanding relativistic fluids with finite spacetime support.
\end{abstract}
\date{\today}
\maketitle
\newpage
\tableofcontents
\section{Introduction and physical motivation}
\label{sec:introduction}
Relativistic hydrodynamics has become a central theoretical framework for describing the collective behavior of strongly interacting matter produced in ultra-relativistic heavy-ion collisions~\cite{Teaney:2009qa,Peralta-Ramos:2010qdp,Gale:2013da,Romatschke:2017ejr,Becattini:2024uha}. In recent years, this framework has been extended to incorporate spin degrees of freedom, motivated by the experimental observation of global and local spin polarization of hadrons~\cite{STAR:2017ckg,STAR:2019erd,ALICE:2019onw,ALICE:2021pzu}. Spin hydrodynamics provides a macroscopic description of spin transport and polarization in systems close to local thermodynamic equilibrium and offers a bridge between microscopic spin–vorticity coupling and experimentally accessible observables~\cite{Becattini:2011zz,Florkowski:2017ruc,Florkowski:2018fap,
Montenegro:2017rbu,Hattori:2019lfp, Fukushima:2020ucl,Hongo:2021ona,Gallegos:2022jow,Das:2022azr,Peng:2021ago,Singh:2025hnb,Cao:2022aku,Weickgenannt:2022zxs,Weickgenannt:2022qvh,Biswas:2023qsw,Becattini:2023ouz,Drogosz:2024gzv,Daher:2025pfq,Armas:2025fvo,Abboud:2025shb,Bhadury:2025wuh}.

It provides a macroscopic description of polarized matter by promoting spin degrees of freedom to independent dynamical variables alongside energy, momentum, and conserved charges. In this approach, the microscopic polarization of constituents is encoded in an antisymmetric tensor field, commonly referred to as the spin potential, which governs the transport and redistribution of angular momentum in the fluid. In stationary situations such as global equilibrium with rigid rotation, this tensor reduces to the familiar thermal vorticity~\cite{Becattini:2007nd,Becattini:2009wh}; however, away from equilibrium, it represents an independent hydrodynamic field whose evolution must be determined dynamically. The resulting framework extends conventional relativistic hydrodynamics by introducing additional equations of motion associated with angular momentum conservation, leading in general to a coupled system that requires numerical or analytic treatment depending on the underlying symmetries.

Analytic flow solutions~\cite{Bjorken:1982qr,Csorgo:2006ax,Bialas:2007iu,Beuf:2008vd,Nagy:2009eq,Gubser:2010ze,Wong:2014sda,Hatta:2014gqa,Shi:2022iyb} play a crucial role in understanding the structure and implications of relativistic (spin) hydrodynamics. Among these, Bjorken flow~\cite{Bjorken:1982qr} provides a simple boost-invariant background, while Gubser flow~\cite{Gubser:2010ze,Gubser:2010ui} extends this picture by including transverse expansion through an $SO(3)$-symmetric embedding into de Sitter space. Recently, a more general class of conformal flows has been identified by Grozdanov in Ref.~\cite{Grozdanov:2025cfx} using a Weyl-rescaled de Sitter construction, labeled by a discrete parameter $\kappa=0,\,\pm 1$, where $\kappa=0$ and $\kappa=+1$ denote the flat (Bjorken) slicing and the spherical (Gubser) slicing of dS$_3\times \mathbb{R}$, respectively. These flows differ by the geometry in the transverse directions and the associated isometry groups\footnote{See Refs.~\cite{Soloviev:2025uig,Martinez:2025jtv} for other interesting works on attractors and the Boltzmann equation using the new analytic solution, respectively.}.

In this work, we focus on the novel $\kappa=-1$ solution, which exhibits $SO(2,1)$ symmetry and corresponds to a hyperbolic slicing of de Sitter space. When mapped back to Minkowski spacetime, this solution describes a transversely expanding fluid with finite spacetime support, bounded by a causal edge inside the future lightcone. This feature sharply distinguishes it from the Gubser $(\kappa=+1)$ solution, which extends to arbitrarily large transverse radius and lacks a natural spacetime boundary. As a result, the $\kappa=-1$ background provides a qualitatively different physical setting, closer to a finite droplet of matter rather than an infinite medium.

We employ the perfect-fluid spin hydrodynamic formulation~\cite{Florkowski:2018fap,Florkowski:2019qdp} and investigate the spacetime evolution of the spin potential on top of a $\kappa=-1$ background. The finite support and enhanced early-time expansion rate of the $\kappa=-1$ flow have important consequences for spin dynamics. The spin evolution equations inherit explicit geometric factors associated with the hyperbolic slicing, leading to stronger early-time dilution and nontrivial mixing between spin components. These effects are absent or parametrically weaker in the $\kappa=+1$ (Gubser) case~\cite{Singh:2020rht}. Studying spin hydrodynamics on the $\kappa=-1$ background therefore allows one to isolate the role of spacetime geometry and causal structure in shaping spin polarization patterns.

The present analysis extends the study of Ref.~\cite{Florkowski:2019qdp}, where the dynamics of spin polarization were examined in a boost-invariant and transversely homogeneous setting. The current study also complements the work done in~\cite{Singh:2020rht,Florkowski:2021wvk} for the Gubser-expanding background and the non-boost-invariant background. However, we find that the new background solution presented in~\cite{Grozdanov:2025cfx} shows non-trivial features of spin degrees of freedom that have the potential for a phenomenological description of high-energy collisions~\cite{ExHIC-P:2020tcv,HADES:2022enx,Sass:2022ucj,BESIII:2025lzd}.

We first solve the perfect-fluid hydrodynamic equations by exploiting conformal symmetry in de Sitter coordinates, which allows one to obtain analytic solutions for the hydrodynamic fields. We then employ the de Groot–van Leeuwen–van Weert (GLW) spin tensor~\cite{DeGroot:1980dk} which, in general, does not comply with conformal symmetry. However, as we are in the perfect-fluid limit, there is no back-reaction from the spin dynamics to the hydrodynamic background evolution. Hence, we relax the constraints associated with special conformal symmetry and retain only cylindrical symmetry together with boost invariance in the spin evolution.

Using the temperature, chemical potential, and flow profiles corresponding to the $\kappa=-1$ background, we obtain the equations of motion for the spin components. These equations exhibit a non-trivial dependence on both de Sitter time and angular variables, as well as a weak sensitivity to the mass of the constituent particles. In the regime of small spin potential considered here, the spin sector does not modify the evolution of the hydrodynamic background fields~\cite{Florkowski:2019qdp}. Consequently, the conformal perfect-fluid solution can be obtained independently, and the spin dynamics may be analyzed subsequently as a probe evolving on a fixed expanding background.

For the special case of massless particles, we identify a class of analytic solutions that typically display a power-law dependence on the local temperature. In addition, we present numerical solutions for systems with finite particle masses in Milne space-time. Our study demonstrates that the $\kappa=-1$ flow constitutes a distinct and valuable analytic laboratory for exploring spin dynamics in relativistic fluids with finite spacetime extent.

The outline of the paper is as follows: in \Cref{sec:perfect}, we provide a recap of spin hydrodynamics for a perfect fluid. Next, we turn our attention to describing the $\kappa=-1$ geometry in the context of high energy collisions. We obtain the evolution of the spin hydrodynamic fields by solving the equations of motion in \Cref{sec:dynamics}
followed by \Cref{sec:summary} where we summarize and describe the physical implications of our work.

\textbf{Conventions.}
We adopt the shorthand notation $a \cdot b\equiv a_\mu b^\mu $ for the scalar product of two four-vectors. The Levi–Civita tensor $\epsilon^{\alpha\beta\gamma\delta}$ is defined with the convention $\epsilon^{0123} = -\epsilon_{0123} = 1$. Natural units are used throughout, with $c = \hbar = k_B = 1$. Antisymmetrization of a rank-two tensor $A_{\mu\nu}$ is defined as
\beq
A_{[\mu\nu]} = \frac{1}{2}\left(A_{\mu\nu} - A_{\nu\mu}\right).
\eeq
Finally, we employ the ``mostly plus'' metric signature.
\section{Perfect fluid spin hydrodynamics in the \texorpdfstring{$\kappa=-1$ geometry}{}}
\label{sec:perfect}
To make the article self-contained, we first summarize the framework of relativistic perfect-fluid hydrodynamics for spin-$\frac{1}{2}$ particles~\cite{Florkowski:2019qdp}. We work in the small-polarization limit, in which the hydrodynamic background (i.e.~the evolution of fluid velocity, temperature, and chemical potential, governed by net baryon number and energy–momentum conservation) decouples from the evolution of the spin degrees of freedom, which follows separately from angular momentum conservation.
\subsection{Perfect-fluid hydrodynamic background}
\label{sec:barcons}
The perfect-fluid relativistic 
hydrodynamic background consists of two conservation laws: the conservation of net baryon number and the conservation of energy-momentum tensor.
Net baryon number conservation is expressed as
\bel{eq:Ncon}
\nabla_\a N^\a(x)  = 0\, ,
\eel
where the net baryon current $N^\a$ is
\bel{eq:N}
N^\a = \cN \, u^\a\, ,
\eel
and the net baryon density (assuming an ideal relativistic gas of classical massive particles) reads 
\bel{eq:calN}
\cN= \frac{2\,T^{3} \, z^2}{\pi^2}  K_{2}\left( z \right) \sh\xi\, .
\eel
In \cref{eq:N}, $u^\a$ is the timelike normalized  fluid four-velocity ($u \cdot u \!=\!-1$) and $\nabla_\alpha$ is the covariant derivative. Further, $\xi$ denotes the ratio of the local baryon chemical potential and the local temperature, $\xi\equiv\mu/T$, and the ratio of particle mass to temperature is $z\equiv m/T$. $K_{n}$ is the $n$-th modified Bessel function of the second kind.
The energy-momentum conservation is 
\bel{eq:Tcon}
\nabla_\a T^{\a\b}(x) = 0\, ,
\eel
where energy-momentum tensor $T^{\a\b}$ takes the following perfect-fluid form
\bel{eq:T}
T^{\a\b} &=& \cE \, u^\a u^\b + \cP \Delta^{\a\b}\, ,
\eel
with $\cE$ and $\cP$ being the energy density and pressure, respectively. For the case of  an ideal relativistic gas of classical massive particles 
\beq
\cE&=& \frac{2\,T^4 \, z^2}{\pi^2} 
 \left[z  K_{1} \left( z \right) + 3 K_{2}\left( z \right) \right]\ch\xi\, ,
\label{eq:calE} \\
\cP&=&   \frac{2\,T^4 \, z^2}{\pi^2}  K_{2}\left( z \right) \ch \xi\, ,
\label{eq:calP}
\eeq where $\Delta^{\a\b} \equiv g^{\a\b} + u^\a u^\b$ is the spatial projection operator orthogonal to $u_\alpha$.
%
\subsection{Conservation of angular momentum}
\label{sec:angcons}
The total angular momentum can be decomposed in terms of orbital angular momentum and spin angular momentum as
\beq
J^{\a,\b\g}=\underbrace{x^{\b}\, T^{\a\g} - x^{\g}\, T^{\a\b}}_{\rm orbital} - \underbrace{S^{\a,\b\g}}_{\rm spin}\, ,
\label{eq:L}
\eeq
where the conservation of total angular momentum leads to
\beq
\nabla_\a J^{\a,\b\g} = 0 \implies \nabla_\a S^{\a,\b\g} = T^{\b\g} - T^{\g\b} \, ,
\eeq
i.e.~the spin tensor is conserved up to the antisymmetric components of $T^{\b\g}$.
As we work with the symmetric energy-momentum tensor, the spin is conserved independently
\beq
\nabla_\a S^{\a,\b\g} = 0 \, .
\label{eq:Scon}
\eeq
In what follows, we will use a particular definition of the spin tensor, known as the de Groot–van Leeuwen–van Weert (GLW) spin tensor, expressed as~\cite{DeGroot:1980dk,Singh:2022uyy}
\beq
S^{\a,\b\g}
&=& S^{\a, \b\g}_{\rm ph}  +  S^{\a, \b\g}_{\Delta \rm GLW}  ,
\label{eq:S}
\eeq
the sum of the phenomenological spin tensor and the correction term arising due to the pseudogauge transformation\footnote{There have been many investigations concerning pseudogauge transformations and ambiguity in the decomposition of total angular momentum into orbital and spin; see Refs.~\cite{Becattini:2011ev,Leader:2013jra,Huang:2020kik,Freese:2021jqs,Singh:2024qvg,Drogosz:2024rbd,Becattini:2025oyi,Becattini:2025twu,Armas:2026bmw,Choi:2026kgf}.},
where
\beq
S^{\a, \b\g}_{\rm ph}&=& \cC \,u^\a \omega^{\b\g}\,, \label{eq:Spheno}\\
S^{\a, \b\g}_{\Delta \rm GLW} 
&=&  \cA \, u^\a u^\d u^{[\b} \omega^{\,\,\g]}_{\d} \label{eq:SDelta}
\, +\cB  \, \Big( 
\Delta^{\a\d}u^{[\b}  \omega^{\g]}_{\HP\d}
+ u^\a \Delta^{\d[\b} \omega^{\g]}_{\HP\d}
+ u^\d \Delta^{\a[\b} \omega^{\g]}_{\HP\d}\Big)\, ,
\eeq
where the thermodynamic coefficients $\cC$, $\cA$, and $\cB$ are defined as
\beq
\cC = \frac{\cP}{4\,T}\, ,
\quad
\cA = 2 \,A -3 \,C\, ,
\quad
\cB = -\frac{\cE+\cP}{2 \,T \, z^2 }\, . \label{eq:thermo_coefficients}
\eeq 
Note that the phenomenological spin tensor, \cref{eq:Spheno}, is well defined in the massless limit, while the full spin tensor, \cref{eq:S}, is not due to explicit mass terms appearing in $S^{\a, \b\g}_{\Delta \rm GLW}$. 

The spin potential $\omega^{\a\b}$ is a rank-two antisymmetric tensor that can be decomposed as
\beq
\omega^{\a\b} &=&  u^\a a^\b - u^\b a^\a  + \epsilon^{\a\b\g\d} u_\g \omega_{\d}\, ,
\label{spinpol}
\eeq
with $a^{\a}$ and $\omega^{\a}$  satisfying the constraints
\beq
a\cdot u = 0\, , \quad \omega \cdot u = 0  \label{ko_ortho}\, .
\eeq
One can express $a_\a$ and $\omega_\a$ in terms of the spin potential as
\beq
a_\a= \omega_{\a\b} u^\b\, , \quad \omega_\a = \half \epsilon_{\a\b\g\d} \omega^{\b\g} u^\d\, . \lab{eq:kappaomega}
\eeq
\subsection{Boost invariant and cylindrically symmetric setup}
\label{sec:implement}
Having recalled the details of spin hydrodynamics, we will discuss in the present section the $\kappa=-1$ geometry presented by Grozdanov in~\cite{Grozdanov:2025cfx}. Consider a system that is boost-invariant and cylindrically-symmetric with respect to the beam ($z$) axis created in head-on heavy-ion high energy collisions. Its dynamics can be described in polar hyperbolic coordinates
$x^\mu = (\tau,r,\phi,\eta)$ where the line element is written as
\beq
ds^2=-d\tau^2+dr^2 + r^2d\phi^2 + \tau^2d\eta^2,
\eeq
where 
\begin{alignat}{3}
    {\text{proper~time:}} \quad  &\tau =\sqrt{t^2 - z^2} \quad && {\rm with} \quad \frac{1}{q}<\tau - r\,,\\
{\text{longitudinal~spacetime~rapidity:}}\quad 
&\eta ={\rm tanh}^{-1}(z/t) \quad && {\rm with } \quad -\infty<\eta < \infty\,,\\
{\text{radial~distance:}}\quad &  r=\sqrt{x^{2}+y^{2}} \quad &&{\rm with} \quad 0<r<\infty\,,\\
{\text{azimuthal~angle:}}\quad &  \phi ={\rm tan}^{-1}(y/x) \quad &&{\rm with} \quad 0\leq \phi < 2\pi.
\end{alignat}
To simplify the calculations in the following sections, we introduce the four-vector basis in the lab frame \begin{align}
u^\mu &= (\ch \vartheta,\sinh \vartheta,0,0)\, , \nonumber \\
R^\mu &= (\sh \vartheta,\ch \vartheta ,0,0)\, , \nonumber \\
\Phi^\mu &= (0,0,1/r,0)\, ,\nonumber \\
Z^\mu &= (0,0,0,1/\tau)\, , 
\label{eq:pmbasis} 
\end{align}
where $\vartheta$ is an arbitrary angle, characterizing the flow velocity. In this form, the inverse metric tensor, \mbox{$g^{\mu\nu}={\rm diag}(-1,1,1/ r^2,1/\tau^2)$}, can be written as
\ba
g^{\mu\nu}=-u^\mu u^\nu+R^\mu R^\nu+\Phi^\mu \Phi^\nu+Z^\mu Z^\nu \, .
\label{eq:pmmetric}
\ea
The four-vector basis are orthogonal to each other and satisfy
\ba
u\cdot u  =-1\,, \quad
R \cdot R =1\, ,\quad  \Phi\cdot \Phi =1\, ,\,\,\quad 
Z\cdot Z =1\, .
\label{eq:relations}
\ea 
One may easily notice that in the local rest frame $u$, $R$, $\Phi$ and $Z$ point, respectively, in the $\tau$, $r$, $\phi$ and $\eta$ directions.

We can decompose the four-vectors  $a^\a$ and $\omega^\a$ further using the four-vector basis \refb{eq:pmbasis} and orthogonality constraints \refb{ko_ortho} as
\beq
a^\a &=&  a_R R^\a + a_\Phi \Phi^\a + a_Z Z^\a, \lab{eq:k_decom}\\
\omega^\a &=&  b_R R^\a + b_\Phi \Phi^\a + b_Z Z^\a, \lab{eq:o_decom}
\eeq
where $a_i(\tau,r)$ and $b_i(\tau,r)$ denote the scalar spin components along the basis vectors.
%
\subsection{Conformal mapping and Weyl transformation}
\label{sec:mapping}
%
For systems that are both boost invariant and cylindrically symmetric, it is possible to construct a nontrivial four-velocity profile in Minkowski space that remains invariant under the conformal symmetry group $SO(2,1)_q \otimes SO(1,1) \otimes \mathbb{Z}_2$~\cite{Grozdanov:2025cfx}. This symmetry group consists of rotations in the $r-\phi$ plane combined with two special conformal transformations characterized by an arbitrary length scale $q$ (forming $SO(2,1)_q$), Lorentz boosts along the spacetime rapidity direction $\eta$ ($SO(1,1)$) and reflections across the $r-\phi$ plane ($\mathbb{Z}_2$).

While determining a flow profile that is invariant under this symmetry group is technically involved when working directly in Minkowski space, the construction becomes considerably more transparent upon mapping the problem to the positively curved spacetime given by the direct product of three-dimensional de Sitter space and a line, dS$_3 \otimes \mathbb{R}$, which we refer to as ``de Sitter space.'' The transformation between these two descriptions is achieved through a conformal mapping implemented via a Weyl rescaling of the spacetime line element 
\beq
ds^2 \to \frac{ds^2}{\Omega^2}=\frac{-d\tau^2+dr^2 + r^2d\phi^2}{\tau^2} + d\eta^2\, ,
\label{eq:deSitter-WR}
\eeq
where $\Omega$ is the conformal weight, $\Omega=\tau$.
After the Weyl transformation, one can transform from polar Milne coordinates $x^\mu=(\tau,\,r,\,\phi,\,\eta)$ to de Sitter  coordinates $\hat{x}^\mu=(\rho,\,\theta,\,\phi,\,\eta)$ via the following coordinate transformations~\cite{Grozdanov:2025cfx}
\ba 
\cosh{\rho} &=&   \frac{1+(q \tau)^2-(q r)^2}{2 q{\tau}}\, ,
\label{eq:desitter1} \\
\tanh{\theta} &=&- \frac{2q r}{1-(q \tau)^2+(q r)^2}\, .
\label{eq:desitter2}
\ea
The resulting line element of ${\rm dS}_3 \otimes \mathbb{R}$ is expressed as
\beq
d\hat{s}^2 = -d\rho^2 + \sinh^2\!\rho \left(d\theta^2 + \sinh^2\!\theta \,d\phi^2\right) + d\eta^2\,,
\label{eq:deSitter-ds}
\eeq
with the metric \mbox{$
\hat{g}_{\mu\nu} = {\rm diag}\left(-1,\, \sinh^2\!\rho,\, \sinh^2\!\rho\,  \sinh^2\!\theta ,\, 1\right)
$}.
Using the transformation rule
\beq
{u}_{\mu}=\tau \frac{\partial \hat{x}^{\nu}}{\partial {x}^{\mu}} \hat{u}_{\nu}\,,
\label{eq:Urel}
\eeq
one can find the four-velocity, see Eq.~(\ref{eq:pmbasis}), to be static in the de Sitter spacetime
\ba 
\hat{u}^\mu &=& (1,\,0,\,0,\,0)\, , \nonumber 
\ea
with the transverse velocity as 
\be 
\frac{u^r}{u^\tau}\equiv v_\perp \equiv\tanh \vartheta(\tau,r)= \left(\frac{2 \,q \tau \,  q r}{(q \tau)^2+(q r)^2-1}\right)\, ,
\label{eq:thetaperp}
\ee 
where the components of $u^\mu$ are
\ba
u^\tau &=& \frac{(q \tau)^2+(q r)^2-1}{\sqrt{1-2 q^2 (\tau^2 + r^2) + q^4 (\tau^2 - r^2)^2}}\,,\\
u^r &=& \frac{2 \,q \tau \,  q r}{\sqrt{1-2 q^2 (\tau^2 + r^2) + q^4 (\tau^2 - r^2)^2}} \,,\\
u^\phi &=& 0 \,,\\
u^\eta &=& 0\,.
\ea
Similarly, the remaining orthogonal basis vectors (\ref{eq:pmbasis}) in the de Sitter space can be written as
\ba 
\hat{R}^\mu &=& (0,\,(\sinh\rho)^{-1},\,0,\,0)\,  , \nonumber \\
\hat{\Phi}^\mu &=& (0,\,0,\,(\sinh\rho \sinh\theta )^{-1},\,0)\, ,   \nonumber \\
\hat{Z}^\mu &=& (0,\,0,\,0,\,1)\, ,
\label{eq:desitter-4vectors}
\ea
where metric $\hat{g}^{\mu\nu}$ reads
\ba
\hat{g}^{\mu\nu}=-\hat{u}^\mu \hat{u}^\nu+\hat{R}^\mu \hat{R}^\nu+\hat{\Phi}^\mu \hat{\Phi}^\nu+\hat{Z}^\mu \hat{Z}^\nu \, ,
\ea
with the determinant of $\hat{g}_{\mu\nu}$ being
\ba
\hat{g} \equiv \det(\hat{g}_{\mu\nu}) = -\sinh^4\!\rho \sinh^2\!\theta\, .  
\label{eq:g}
\ea
Finally, we note the non-vanishing components of the Christoffel symbol

\begin{alignat}{3}
\Gamma^\rho_{\theta\theta}&=\sinh\rho\cosh\rho\, ,   \quad 
&&\Gamma^\rho_{\phi\phi}=\sinh^2\theta\sinh\rho\cosh\rho\, ,  \quad  
&\Gamma^\theta_{\rho\theta}=\Gamma^\theta_{\theta\rho}=\coth\rho\, ,   \\
\Gamma^\theta_{\phi\phi}&=-\sinh\theta\cosh\theta\, ,   \quad
&&\Gamma^\phi_{\rho\phi}=\Gamma^\phi_{\phi\rho}=\coth\rho\, ,   \quad 
&\Gamma^\phi_{\theta\phi}=\Gamma^\phi_{\phi\theta}=\coth\theta\, .   
\end{alignat}
\section{Evolution in de Sitter coordinates}
\label{sec:dynamics}
%
In this section, we investigate the evolution of spin degrees of freedom on a conformally expanding perfect-fluid background in de Sitter spacetime, employing the newly identified maximally symmetric flow, which is the hyperbolic slicing of de Sitter space $(\kappa=-1)$~\cite{Grozdanov:2025cfx}.

The corresponding solutions in Minkowski spacetime are then obtained by mapping the results back through Weyl and coordinate transformations~\cite{Singh:2020rht}. Although the ideal hydrodynamic solution formally develops singular behavior at the causal edge, the present analysis focuses on the interior region of the droplet where gradients remain finite and the perfect-fluid description is expected to be reliable.
\subsection{Equation of state}
To use the conformally-expanding flow, conformal symmetry needs to be respected for vanishing particle mass. Therefore, we write the energy density \rfn{eq:calE}, pressure \rfn{eq:calP} and net baryon density \rfn{eq:calN} in the massless limit as
\beq
\hat{\cE}&=&\frac{12\,{\hat{T}}^4}{\pi^2} \ch{\xi}\,,  
\label{eq:GE} \\
\hat{\cP}&=&\frac{4\,{\hat{T}}^4}{\pi^2} \ch{\xi}\,,  
\label{eq:GP} \\
\hat{\cN}&=&\frac{4\,{\hat{T}}^3}{\pi^2} \sh{\xi}\,,  
\label{eq:GN}
\eeq
respectively, which satisfies $\hat{\cE}=3\hat{\cP}$.

As mentioned previously, the GLW expression \cref{eq:S} for the spin tensor employed in this work is not strictly well defined in the massless limit. In the fully coupled treatment of \cref{eq:Ncon,eq:Tcon,eq:Scon}, the GLW term would generically induce a breaking of conformal symmetry and, consequently, of the associated flow invariance. In the specific setup considered here, however, this complication does not arise because the spin dynamics governed by \rfn{eq:Scon} is incorporated only perturbatively~\cite{Florkowski:2019qdp} and, therefore, decouples from the perfect-fluid background. As a result, the background equations of motion \rfn{eq:Ncon} and \rfn{eq:Tcon} can be solved independently, followed by a separate determination of the spin sector described by \rfn{eq:Scon}.

As the evolution of spin degrees of freedom does not feed back into the background, i.e.~the spin potential does not appear in \cref{eq:Ncon,eq:Tcon}, the breaking of conformal symmetry that the spin dynamics induces does not invalidate the underlying flow invariance at this order. 
A closely related situation has also been discussed in other recent studies; see, for example, Refs.~\cite{Du:2020bxp,Singh:2020rht}.
It is therefore justified, despite this limitation, to analyze the spin dynamics directly in de Sitter coordinates, which we adopt for convenience. By the same reasoning, we retain finite particle masses in the expressions for ${A}$, ${B}$, and ${C}$ that define the spin tensor in \rfn{eq:S}.
%
\subsection{Perfect-fluid hydrodynamic evolution}
\label{ss:pfb}
%
The conservation equation for net baryon current \cref{eq:Ncon} can be written as
\beq
\hat{u}^{\a}\p_{\a}\hat{\cN}+\hat{\cN}\p_{\a}\hat{u}^{\a} +\hat{\cN} \hat{u}^{\a} \frac{\p_\a \sqrt{-\hat{g}}}{\sqrt{-\hat{g}}}=0\, ,
\label{eq:chargedS}
\eeq
with $\hat{g}$ being the determinant of the de Sitter metric~(\ref{eq:g}). Equation~\eqref{eq:chargedS} can be re-written in terms of the expansion scalar
\beq
\hat{\theta}\equiv \nabla_\mu \hat{u}^\mu \equiv 2\coth{\rho}\,,
\label{eq:expansion_scalar}
\eeq
as
\beq
\p_\rho\hat{\cN}+\hat{\cN} \hat{\theta}=0\,. 
\lab{eq:charge}
\eeq
Note that the expansion scalar \eqref{eq:expansion_scalar} is a hyperbolic cotangent function of the de Sitter time $\rho$, where the early-time (small $\rho$) expansion rate is parametrically larger because $\hat{\theta}$ diverges as $\rho \rightarrow 0$. That means stronger dilution of densities and stronger ``Hubble-like friction'' terms in any comoving evolution equation.

Taking the timelike projection along the four-velocity, $\hat{u}_{\b}$, of the conservation of the ideal energy momentum tensor \cref{eq:Tcon} in the de Sitter geometry leads to
\beq
\p_\rho\hat{\cE} + \left(\hat{\cE}+\hat{\cP}\right) \hat{\theta}=0\,.
\lab{eq:en}
\eeq
It is easy to verify that the other projections of \cref{eq:Tcon} are satisfied trivially. Due to the de Sitter symmetry, the system's dynamics only depend on the de Sitter time $\rho$.

Equations \cref{eq:charge,eq:en} can be solved analytically~\cite{Grozdanov:2025cfx}
\beq
\hat{\cE}&=& \hat{\cE}_0 \sinh^{-8/3}{\rho}, 
\label{eq:GESol} \\
\hat{\cN}&=& \hat{\cN}_0 \sinh^{-2}{\rho}, 
\label{eq:GNSol}
\eeq
with $\hat{\cE}_0\equiv\hat{\cE}(\rho_0)$ and $\hat{\cN}_0\equiv\hat{\cN}(\rho_0)$ are constants of integration at the initial de Sitter time $\rho_0$. 
The evolution for temperature and baryon chemical potential can also be obtained easily using \cref{eq:GE,eq:GN}
\ba
\label{eq:GTSol}
\hat{T}&=& \hat{T}_0\sinh^{-2/3}{\rho},  \\
\label{eq:GMuSol}
\hat{\mu}&=& \hat{\mu}_0\sinh^{-2/3}{\rho},
\ea
keeping the ratio $\hat{\xi}\equiv\hat{\mu}/\hat{T}$ constant. Here, $\hat{T}_0\equiv\hat{T}(\rho_0)$ and $\hat{\mu}_0\equiv\hat{\mu}(\rho_0)$ are integration constants. 

\begin{figure}[ht!]
\begin{center}
\includegraphics[width=8.1cm]{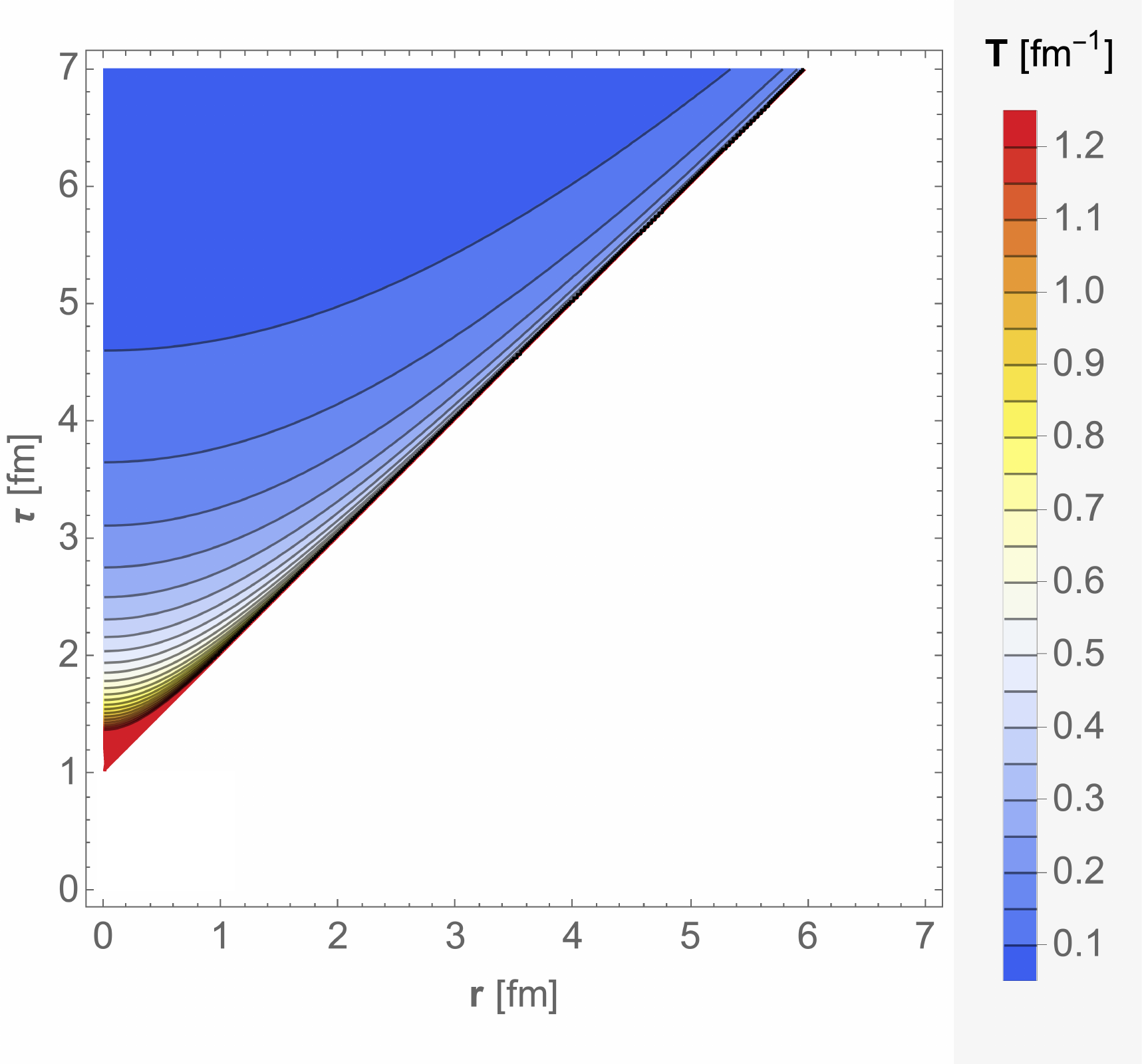}
\end{center}
\caption{The temperature evolution as a function of proper-time $\tau$ and radial distance $r$. }
\label{fig:TandU}
\end{figure}

The temperature evolution as a function of proper-time and radial distance is shown in \Cref{fig:TandU}, which localizes along the lightcone~\cite{Grozdanov:2025cfx}. We followed Ref.~\cite{Singh:2020rht} for the initial temperature profile, where $\hat{T}_0\equiv\hat{T}(\rho_0)=1.2$ at $\rho_0=0.5$, which, in turn, means  that $q\!=\!1 \,\rm{fm^{-1}}$ gives \mbox{$T(\tau_0=1 \,{\rm fm}, r=0)=1.2 \,{\rm fm^{-1}}$}. 

In contrast to the $\kappa=1$ (Gubser) background, which is ``globally smooth'' in the transverse plane and has closed-slice geometry, the $\kappa=-1$ background corresponds to a negatively curved slicing, which geometrically prefers ``expansion towards a causal edge'' rather than an everywhere-smooth profile. One can readily notice that $\kappa=1$ is a radially expanding solution with infinite transverse extent, whereas $\kappa=-1$ is a radially expanding droplet with finite support, ending at a causal edge where the ideal fields become singular (and, with dissipation, one expects a breakdown near the edge)~\cite{Singh:2020rht,Grozdanov:2025cfx}.
This difference will affect the spin dynamics, as we shall see below, because the spin evolution equations inherit explicit functions of the slicing through the ${\rm dS}_3$ metric, which change the qualitative behavior.
\subsection{Spin evolution}
\label{subsec:spin_dynamics}
In this section, we derive the equations of motion for the spin degrees of freedom characterized by the spin potential $\omega_{\a\b}$ on top of a maximally-symmetric perfect-fluid background presented in the previous section using expressions for energy density \rfn{eq:calE} and pressure \rfn{eq:calP} for finite particle masses. 
We also assume that the spin components depend on both $\rho$ and $\theta$ coordinates and derive the equations in de Sitter spacetime. 

Using the decompositions \cref{eq:k_decom} and \cref{eq:o_decom},
we then project the conservation of the spin current \cref{eq:Scon} onto the basis components $\hat{u}_\b \hat{R}_\g$, $\hat{u}_\b \hat{\Phi}_\g$, $\hat{u}_\b \hat{Z}_\g$,  $\hat{\Phi}_\b \hat{Z}_\g$, $\hat{R}_\b \hat{Z}_\g$ and $\hat{R}_\b \hat{\Phi}_\g$. Projecting \cref{eq:Scon} onto the orthonormal basis $\{\hat{u}^\mu, \hat{R}^\mu, \hat{\Phi}^\mu, \hat{Z}^\mu\}$, we obtain six coupled evolution equations for the independent components of the spin potential, which are given by
\begin{align}
\hat{\cB}~\dot{\hat{a}}_R &= -\hat{a}_R \left[\dot{\hat{\cB}} + \frac{5}{2} \hat{\cB} \coth{\rho}\right] \, ,  \label{eq:ar}\\
\hat{\cB}~\dot{\hat{a}}_\Phi + \frac{\hat{\cB}}{2} \sinh{\rho} \sinh{\theta} \,{\hat{b}}^\prime_Z &= -\hat{a}_\Phi \left[\dot{\hat{\cB}} + \frac{5}{2} \hat{\cB} \coth{\rho}\right]- \hat{b}_Z\frac{\hat{\cB}}{2}  \sinh{\rho} \cosh{\theta} \, , \label{eq:apbz1}\\
\hat{\cB}~\dot{\hat{a}}_Z - \frac{\hat{\cB}}{2} \sinh{\rho} \sinh{\theta} \,{\hat{b}}^\prime_\Phi &= -\hat{a}_Z \left[\dot{\hat{\cB}} +\, 3 \,\hat{\cB} \coth{\rho}\right]+ \hat{b}_\Phi\,\hat{\cB}\sinh{\rho}\cosh{\theta}  \, , \label{eq:azbp1}\\
\left(\hat{\cB}-\hat{\cC}\right)~\dot{\hat{b}}_R &= -\hat{b}_R \left[\left(\dot{\hat{\cB}}-\dot{\hat{\cC}}\right) + \left(\frac{9\hat{\cB}}{2}-4\hat{\cC}\right) \coth{\rho}\right] \, , \label{eq:br}\\
\left(\hat{\cB}-\hat{\cC}\right) \dot{\hat{b}}_\Phi -\frac{\hat{\cB}(\csch{\rho})^3}{2\sinh{\theta}} {\hat{a}}^\prime_Z &= -\hat{b}_\Phi \left[\left(\dot{\hat{\cB}}-\dot{\hat{\cC}}\right) + \left(\frac{9\hat{\cB}}{2}-4\hat{\cC}\right) \coth{\rho}\right] \, , \label{eq:azbp2}\\
\left(\hat{\cB}-\hat{\cC}\right) \dot{\hat{b}}_Z +\frac{\hat{\cB}(\csch{\rho})^3}{2\sinh{\theta}} {\hat{a}}^\prime_\Phi &= -\hat{b}_Z \left[\left(\dot{\hat{\cB}}-\dot{\hat{\cC}}\right) + \left(5\hat{\cB}-4\hat{\cC}\right) \coth{\rho}\right] - \frac{\hat{\cB}\coth{\theta}(\csch{\rho})^3}{2\sinh{\theta}} \hat{a}_\Phi \, ,\nn\\
\label{eq:apbz2}
 \end{align}
where $\dot{(\HP)}$ $\equiv$ $\p_\rho$, ${(\HP)}^\prime$ $\equiv$ $\p_\theta$. We note that the hyperbolic slicing induces explicit $\coth{\rho}$, $\csch{\rho}$ and $\theta$-dependent mixing terms absent in the Bjorken case~\cite{Florkowski:2019qdp}.
The thermodynamic coefficients (in de Sitter coordinates) $\hat{\cB}$ and $\hat{\cC}$ are defined in \cref{eq:thermo_coefficients}.
We have also derived the equations of motion concentrating solely on the phenomenological spin tensor \cref{eq:Spheno}. However, due to conformal symmetry, the solutions for spin components exhibit similar behavior, see Appendix~\ref{app:pheno}.
\subsubsection{Special analytic solutions in the massless limit}
\label{sss:massless}
We observe that, spin components along $R^\mu$ evolve independently and do not couple with other spin components, unlike in the related studies~\cite{Florkowski:2019qdp,Florkowski:2021wvk}. However, the coupling happens between $\hat{a}_\Phi$ ($\hat{b}_\Phi$) and $\hat{b}_Z$ ($\hat{a}_Z$). This is a consequence of the conformal symmetry~\cite{Singh:2020rht}.

We can find analytic solutions in the massless limit for \cref{eq:ar,eq:br}
\begin{equation}
\label{eq:ar0}
\hat{a}_R= \hat{a}_R^0 \sinh^{5/6}{\rho}, \quad
\hat{b}_R= \hat{b}_R^0\sinh^{-7/6}{\rho}\,,
\end{equation}
with the superscript zero denoting the integration constant, e.g.~$\hat{a}_R^0\equiv\hat{a}_R(\rho_0)$.
We note that the convex dependence of $\hat{a}_R$ on de Sitter time $\rho$ is in contrast to 
 $\hat{b}_R$, which exhibits a characteristic concave profile, qualitatively resembling the behavior of the temperature and the baryon chemical potential. 

The evolution of the spin components $\hat{b}_\Phi$ and $\hat{a}_Z$ in \cref{eq:azbp1,eq:azbp2} is more involved and displays several distinctive features. 
Let us first consider the case where the $\hat{b}_\Phi$ component is initially negligible. In this situation, the evolution equation for $\hat{a}_Z$ simplifies considerably and admits an analytic solution that is independent of the angular variable $\theta$. One finds 
$\hat{a}_Z|_{\hat{b}_\Phi=0}=\hat{a}_Z^0 \sinh^{1/3}{\rho}\, 
$.
Using the background temperature profile, this solution implies $\hat{a}_Z(\rho)\sim
{\hat{T}(\rho)}^{-1/2}$, indicating that the spin dynamics follows the dilution pattern set by the hyperbolic expansion. In contrast to the Gubser case~\cite{Singh:2020rht}, this behavior reflects the localization of spin dynamics toward the lightcone region of a finite droplet rather than a homogeneous longitudinal expansion.

When $\hat{b}_\Phi$ is non-zero, the evolution of $\hat{a}_Z$ and $\hat{b}_\Phi$ becomes coupled through angular-dependent mixing terms. However, the numerically observed weak dependence of $\hat{a}_Z$ on $\theta$ allows one to neglect the corresponding derivative term in \cref{eq:azbp2} as a controlled approximation. Under this assumption, the equation governing $\hat{b}_\Phi$ admits an approximate analytic solution
\ba
\label{eq:bP}
\hat{b}_\Phi&\approx& \hat{b}_\Phi^0\sinh^{-7/6}{\rho}\,.
\ea
The validity of this approximation is confirmed by exact numerical solutions, which show that the residual angular dependence of $\hat{b}_\Phi$ remains weak throughout the evolution. The scaling behavior of $\hat{b}_\Phi$ closely parallels that of the radial component $\hat{b}_R$, reflecting the common geometric origin of their dilution factors in the hyperbolic background.
Note, the angular dependence enters the evolution equations exclusively through mixing terms suppressed by inverse powers of $\sinh{\rho}$, which rapidly decrease away from the early-time region, explaining the numerically observed mild $\theta$-dependence of the solutions, see \cref{eq:azbp2}.

Closed-form analytic solutions for the remaining components $\hat{a}_\Phi$ and $\hat{b}_Z$ are not available in the general case due to the fully coupled structure of \cref{eq:apbz1,eq:apbz2}. Nevertheless, the particular form of the angular mixing terms allows one to identify a special class of solutions by imposing the vanishing of the explicit $\theta$-dependent contributions. Under this condition, the equations decouple and yield
\ba
\label{eq:aP}
\hat{a}_\Phi&\approx& \hat{a}_\Phi^0   \frac{\sinh^{5/6}{\rho}}{\sinh \theta}\,, \qquad
\hat{b}_Z \approx    \frac{\hat{b}_Z^0}{\sinh^{5/3}{\rho}\sinh\theta}\!, 
\ea
where  $\hat{a}_\Phi^0\equiv\hat{a}_\Phi(\rho_0)$ and $\hat{b}_Z^0\equiv\hat{b}_Z(\rho_0)$. These solutions illustrate how the hyperbolic slicing naturally induces both radial dilution and angular localization of spin dynamics, with $\hat{a}_\Phi$ exhibiting a scaling behavior similar to that of $\hat{a}_R$. It is important to highlight that these analytic expressions should be understood as special solutions illustrating the dominant geometric scaling of the spin components in the massless limit; generic initial conditions require solving the full coupled system numerically.

It is interesting to observe that the above analytic solutions are structurally similar to the $\kappa=+1$ (Gubser) case presented in~\cite{Singh:2020rht} due to the underlying geometry, yet differ precisely due to the hyperbolic slicing of de Sitter space.
\subsubsection{Numerical analysis}
\label{sss:Results}
Let us now move on to the numerical analysis of the equations of motion for the spin components. For $\kappa=-1$, the mapping back to Minkowski space restricts the physical domain to a finite region inside the future lightcone, implying the existence of a causal edge where gradients become large. We remind the reader that the de Sitter quantities are hatted, while the  Minkowski quantities are not.
\begin{figure}[ht!]
\begin{center}
\includegraphics[width=8.1cm]{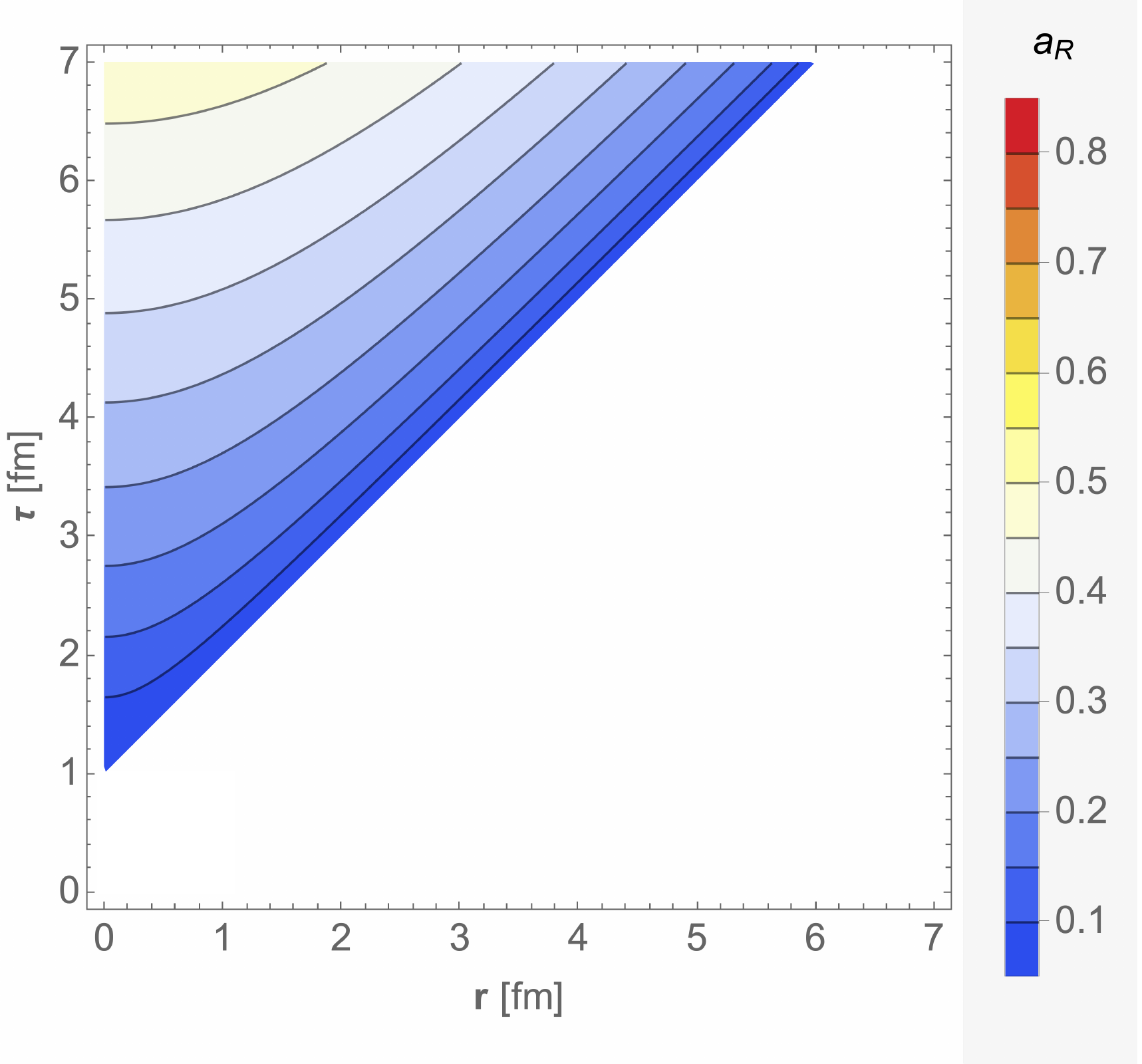}
\includegraphics[width=8.1cm]{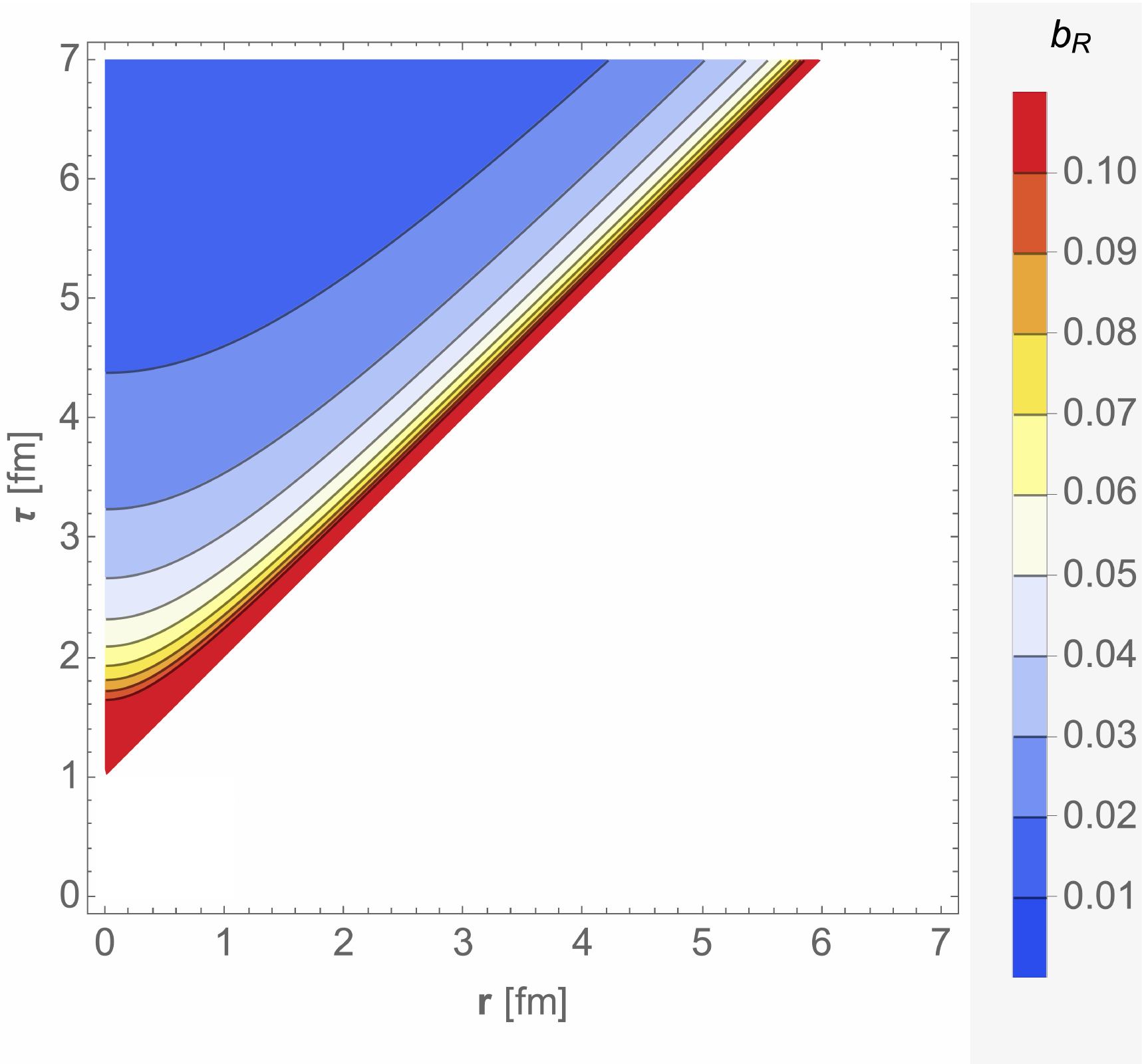}
\end{center}
\caption{The evolution of $a_R$ and $b_R$ spin components as a function of proper-time $\tau$ and radial distance $r$.}
\label{fig:arbr}
\end{figure}

\Cref{fig:arbr} displays the exact numerical solutions of \cref{eq:ar,eq:br} for the radial spin components $a_R$ and $b_R$ mapped to Milne spacetime. The initial conditions for the $\hat{a}_R$ and $\hat{b}_R$ components are fixed with $\hat{a}_R^0 = 0.1$ and $\hat{b}_R^0 = 0.1$, which correspond to
\beq
a_R(\tau_0 = 1 \,{\rm fm}, r = 0) = b_R(\tau_0 = 1 \,{\rm fm}, r = 0) = 0.1 .
\eeq
The particle mass is chosen as $m = 0.5\,\hat{T}_0$.
We observe that the spacetime evolution of the $a_R$ and $b_R$ components differs qualitatively near the initial de Sitter time $\rho_0$: $a_R$ develops a minimum, whereas $b_R$ exhibits a maximum at the initial de Sitter time $\rho_0$. This behavior is primarily driven by the geometric expansion encoded in the $\coth{\rho}$ terms characteristic of the $\kappa=-1$ background.

Comparison with the massless limit confirms that finite-mass effects remain subleading for $m/T < 1$, indicating that the dominant features of the evolution are controlled by the underlying hyperbolic geometry rather than by explicit mass scales. We observe that, unlike the Gubser background, the $\kappa=-1$ geometry introduces a physical boundary in Minkowski space, making the spin evolution sensitive to near-edge regions even in the perfect-fluid limit.

\begin{figure}[hbp]
\begin{center}
\includegraphics[width=8.1cm]{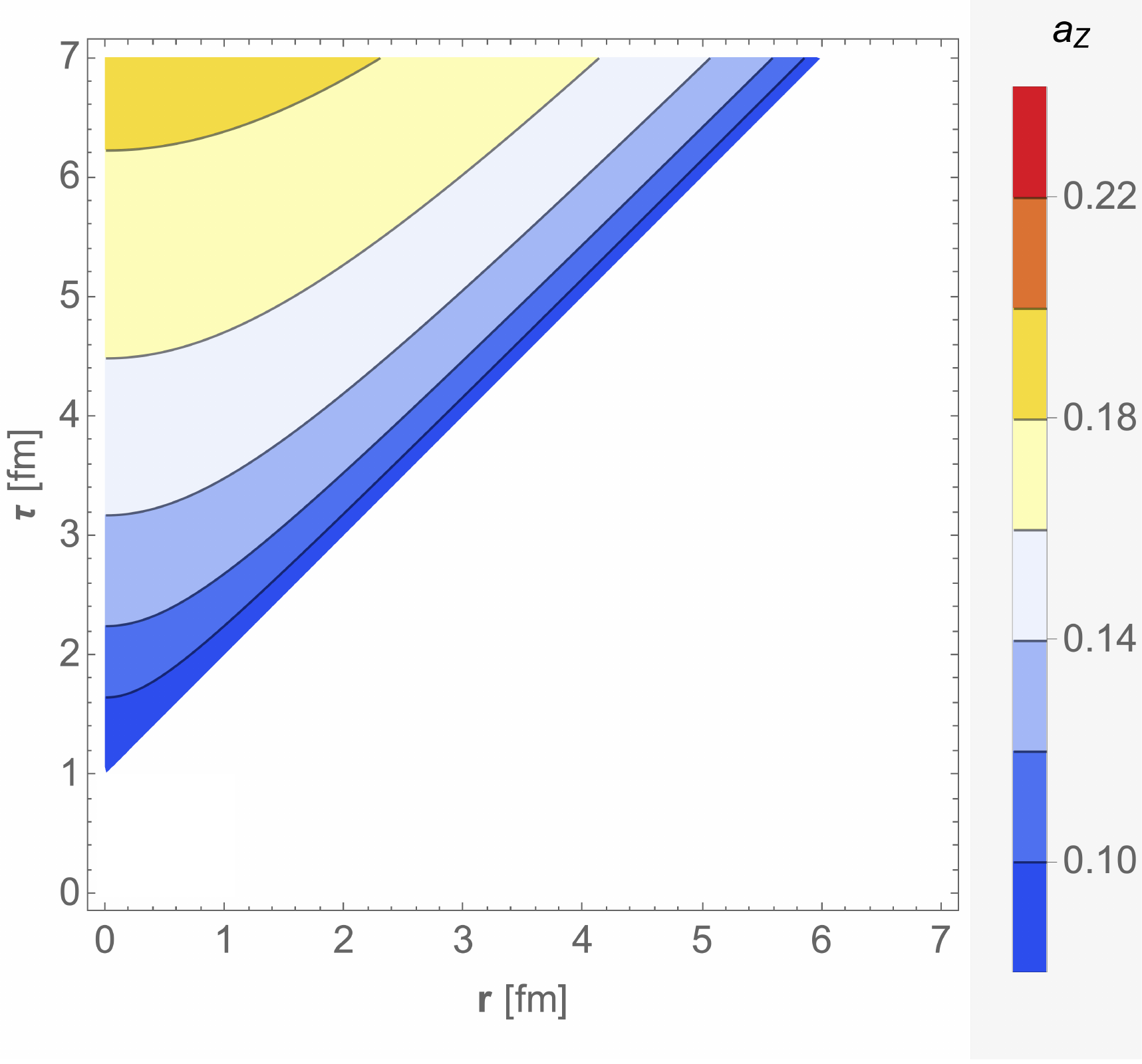}
\includegraphics[width=8.1cm]{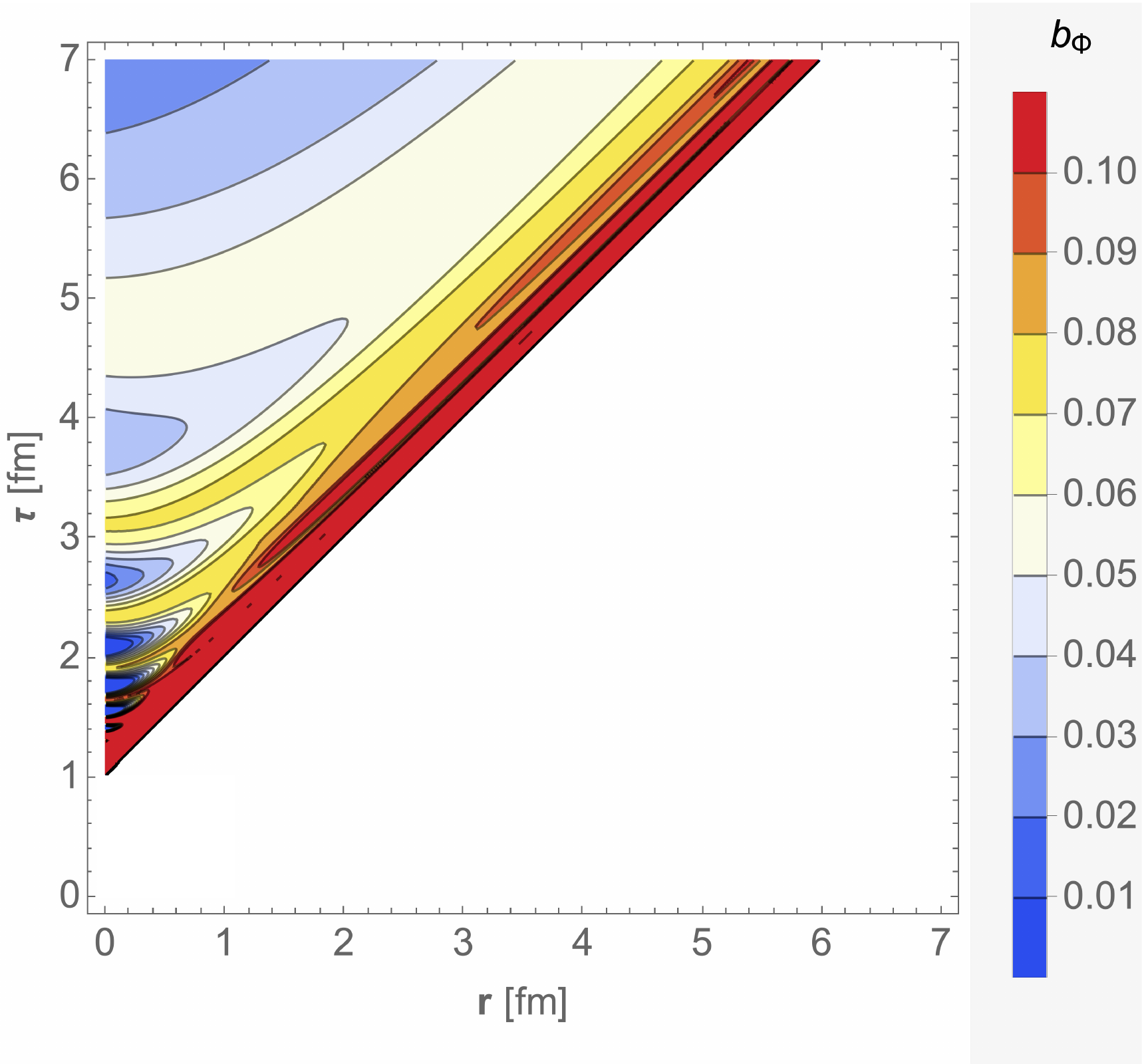}
\end{center}
\caption{The evolution of $a_Z$ and $b_\Phi$ spin components as a function of proper-time $\tau$ and radial distance $r$. }
\label{fig:azbp}
\end{figure}

\Cref{fig:azbp} shows the numerical solutions of \cref{eq:azbp1,eq:azbp2} for the coupled spin components $a_Z$ and $b_\Phi$, obtained with initial values $\hat{a}_Z^0 = \hat{b}_\Phi^0 = 0.1$. Both components display a weak dependence on the angular coordinate $\theta$, reflecting the presence of mixing terms in the evolution equations. In contrast to the Gubser-expanding background, the $\kappa=-1$ flow corresponds to a finite, causally bounded droplet in Minkowski space. As a result, the spin dynamics is naturally influenced by the vicinity of the causal edge, where gradients are enhanced. This manifests most clearly in the behavior of the $b_\Phi$ component, which develops oscillatory features associated with the finite spacetime support of the background. Such edge-driven effects are absent in the Gubser case~\cite{Singh:2020rht}, where the transverse profile is globally smooth and the spin evolution is governed by more uniformly distributed gradients.

The $\kappa=-1$ background describes a finite, causally bounded droplet in Minkowski space rather than an everywhere defined transverse profile, and its de Sitter time expansion rate is controlled by $\coth{\rho}$, leading to parametrically stronger early-time dilution than the bounded $\tanh{\rho}$ behavior of the Gubser background~\cite{Gubser:2010ze}. These geometric differences enter the spin sector directly through the coefficients of the spin potential evolution equations. The presence of $\coth{\rho}$ and $\csch{\rho}$ factors enhances early-time damping and transient mixing between spin components, and after mapping back to Milne coordinates, the dynamics become naturally localized toward the causal region of the droplet.

\begin{figure}[ht!]
\begin{center}
\includegraphics[width=8.1cm]{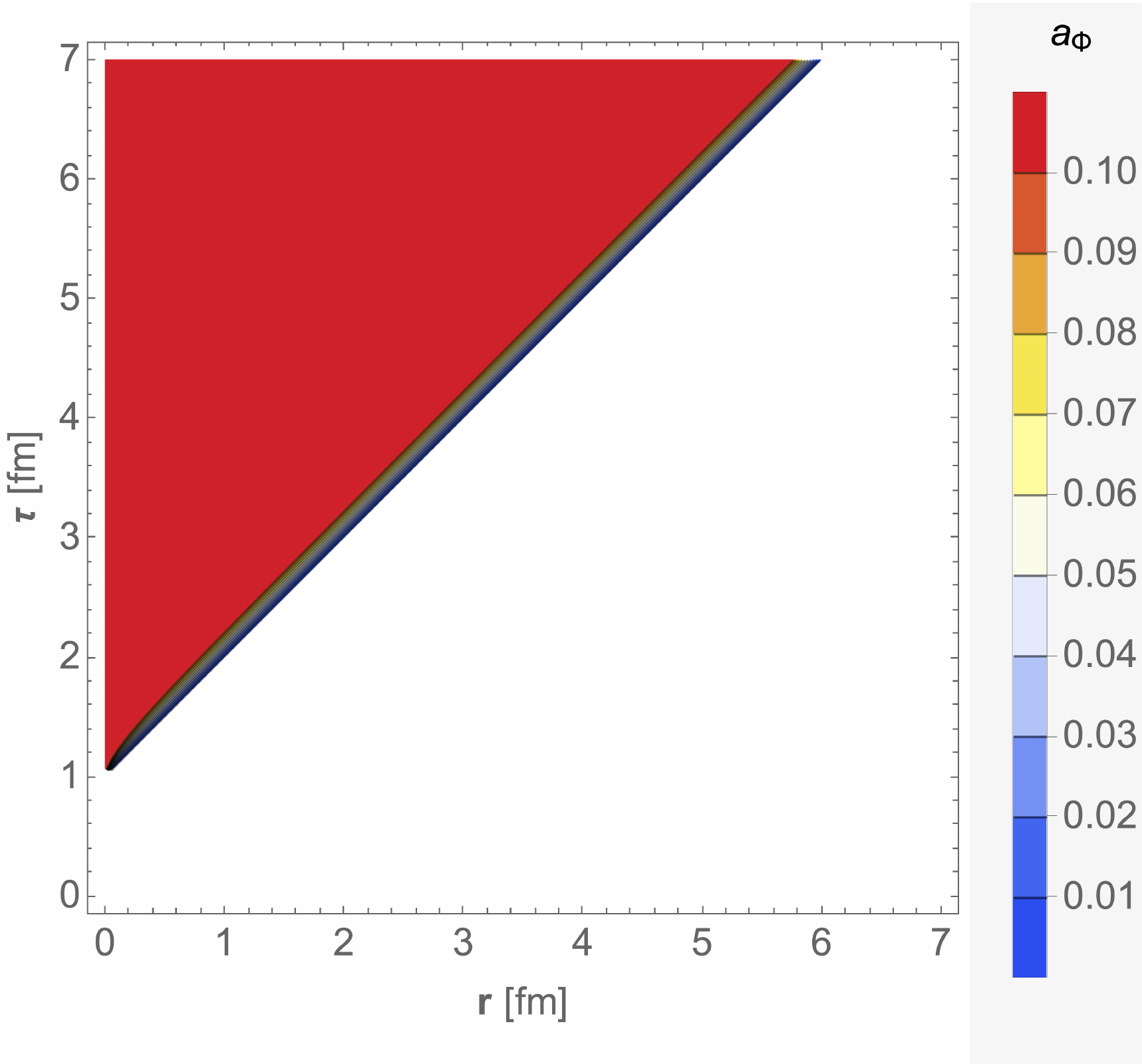}
\includegraphics[width=8.1cm]{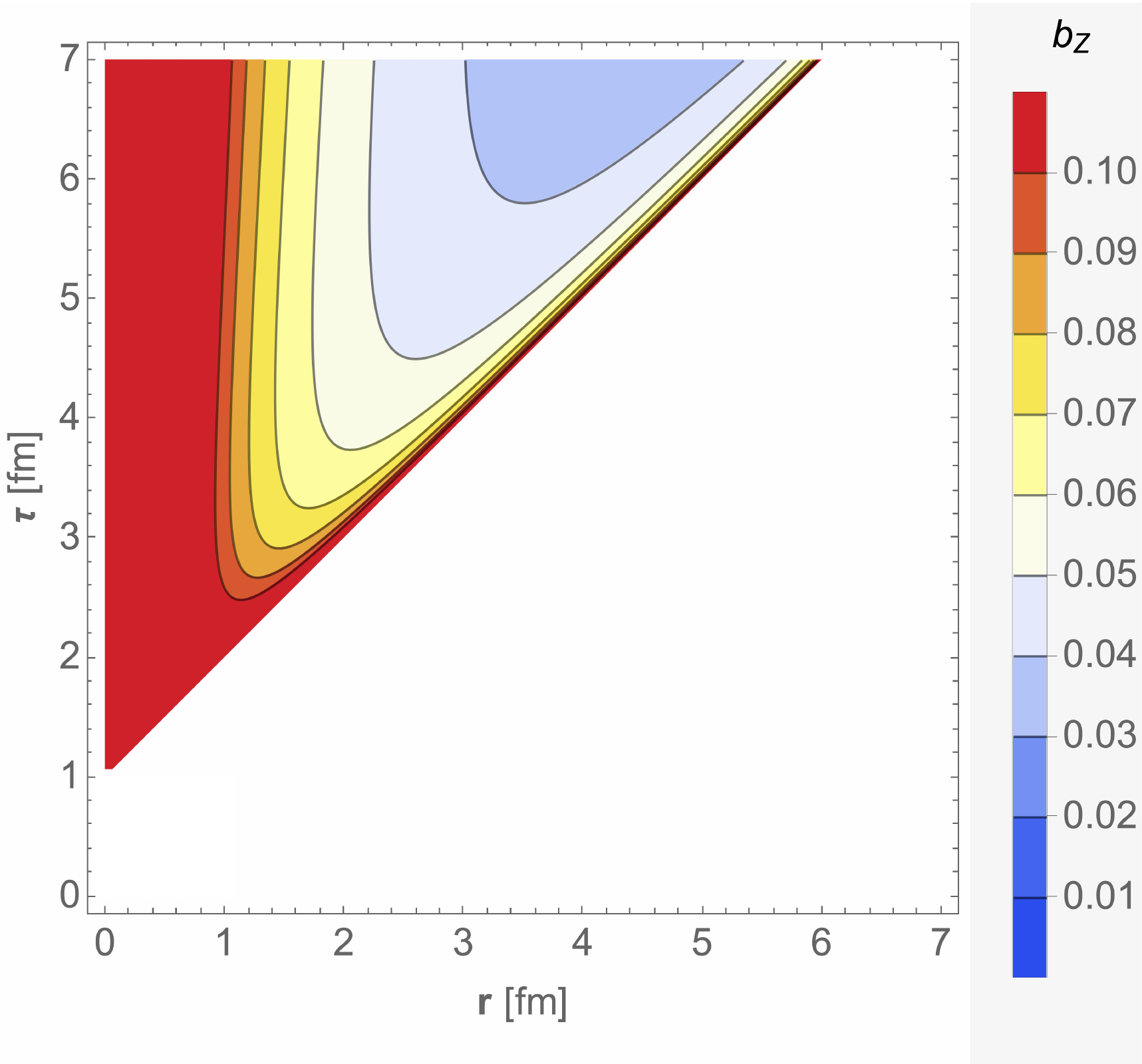}
\end{center}
\caption{The evolution of $a_\Phi$ and $b_Z$ spin components as a function of proper-time $\tau$ and radial distance $r$.}
\label{fig:bzap}
\end{figure}

A comparison with the massless limit confirms that the solutions are only weakly sensitive to the particle mass. In addition, we find that increasing the mass generally enhances the magnitude of the spin components within the region under consideration. Similar qualitative behavior is observed for the remaining $a_\Phi$ and $b_Z$ components, as illustrated in \Cref{fig:bzap}.
\subsubsection{Origin of the oscillatory behavior of \texorpdfstring{$b_\Phi$}{} near \texorpdfstring{$r \rightarrow 0$}{}}
\begin{figure}[ht!]
\begin{center}
\includegraphics[width=0.5\linewidth]{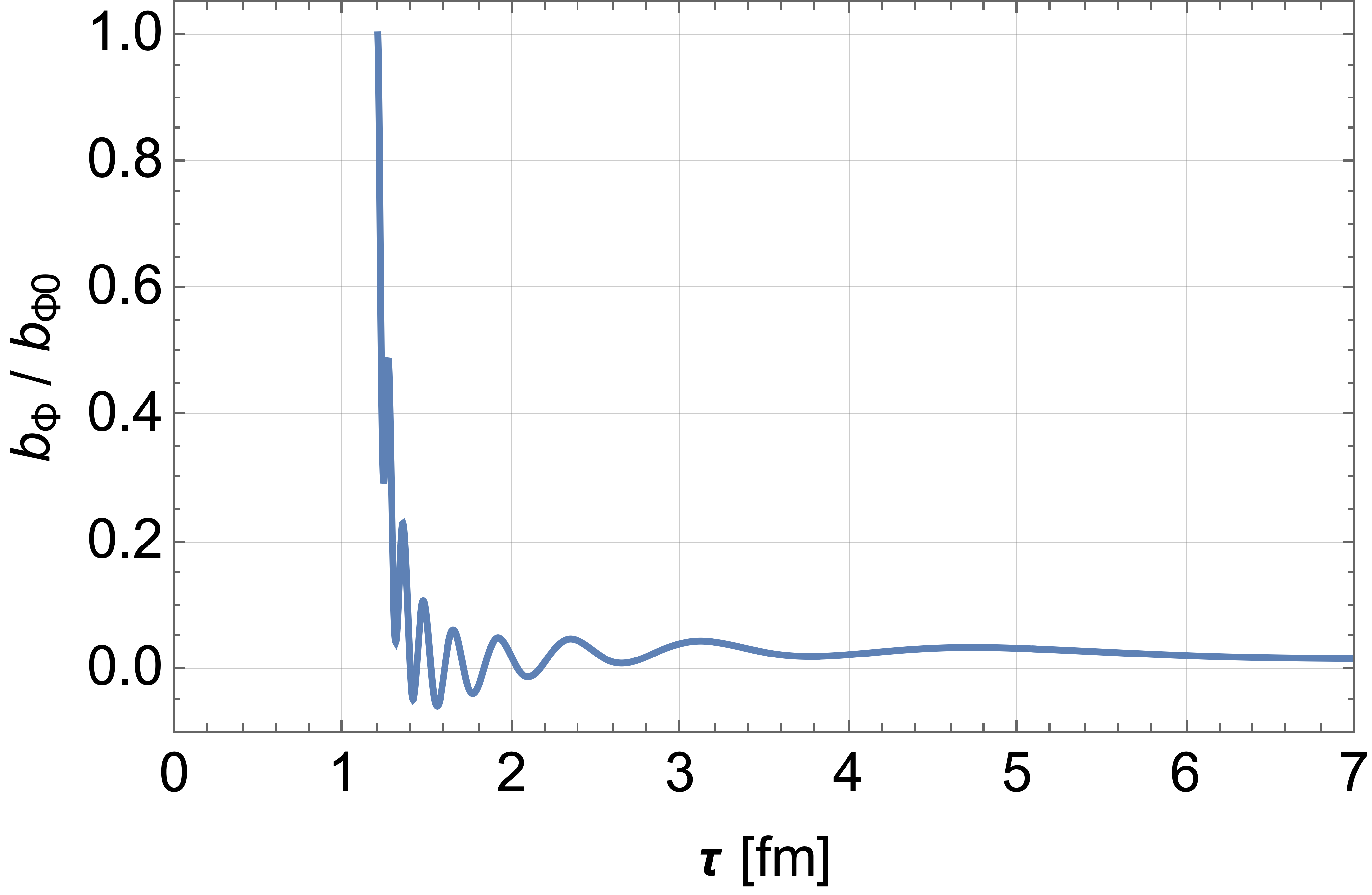}
\end{center}
\caption{Oscillatory behavior of normalized $b_\Phi$ at $r= 0$.}
\label{fig:bposc}
\end{figure}

Although several spin components are coupled through geometric factors generated by hyperbolic slicing, only the azimuthal component $b_\Phi$ develops oscillatory behavior, see \Cref{fig:bposc}. The reason is not the mere presence of coupling, but the specific structure of that coupling. The evolution of $b_\Phi$ is driven exclusively by derivative-type angular mixing with the longitudinal component $a_Z$, such that its dynamics are governed by a competition between early-time geometric mixing and expansion-induced damping, see~\cref{eq:azbp2}. In the absence of additional non-derivative angular source terms, this competition can temporarily produce under-damped, wave-like behavior, which manifests as transient oscillations.

By contrast, the evolution of $b_Z$ in~\cref{eq:apbz2}, although also coupled to another spin component, differs qualitatively. In addition to derivative mixing with $a_\Phi$, the $b_Z$ equation contains an explicit non-derivative term proportional to $a_\Phi\coth{\theta}$. This term, which becomes large near the symmetry axis, enforces the regularity of azimuthal spin modes and effectively acts as an angular restoring or forcing contribution rather than a propagating one. As a result, the $(a_\Phi,b_Z)$ sector is dynamically more constrained, and the evolution of $b_Z$ is dominated by monotonic relaxation and dilution instead of oscillatory motion.

The remaining spin components exhibit even simpler behavior. The radial components $a_R$ and $b_R$ are fully decoupled from angular dynamics by symmetry and evolve solely through geometric dilution, precluding any oscillatory response. The components $a_Z$ and $a_\Phi$ enter the evolution equations only at first order in de Sitter time and act primarily as sources for the $b$ components, rather than as independent propagating modes. Consequently, they do not develop oscillations themselves.

In summary, these features imply that the oscillatory behavior of $b_\Phi$ is a geometry-driven and symmetry-based effect, specific to the azimuthal spin degree of freedom in finite, hyperbolically expanding systems. It reflects the unique role of $b_\Phi$ as the only component that probes angular spin transport without being constrained by additional regularity-enforcing source terms. This explains why oscillations are present in $b_\Phi$ but absent in $b_Z$ and in all other spin components, despite the presence of geometric mixing in multiple sectors.
\section{Summary and outlook}
\label{sec:summary}
In this work, we have investigated relativistic spin hydrodynamics on the hyperbolic $\kappa=-1$ flow background \cite{Grozdanov:2025cfx}. Starting from the perfect-fluid spin hydrodynamic formulation, we derived the complete set of coupled evolution equations for the spin potential on this background and solved them numerically for both massless and massive particles. The $\kappa=-1$ flow differs fundamentally from the Gubser $(\kappa=+1)$ solutions in that it describes a transversely expanding fluid with finite spacetime support and a well defined causal edge.

We showed that the geometric properties of the $\kappa=-1$ background lead to a parametrically stronger early-time expansion and enhanced mixing between spin components. When mapped back to Minkowski space, these features manifest as a localization of spin dynamics toward the interior of the lightcone, in contrast to the more uniformly distributed transverse dynamics found in the Gubser case. Finite particle mass modifies the magnitude and transient behavior of the spin but does not alter the qualitative structure dictated by the background geometry. The key distinction from previously studied Gubser-expanding backgrounds~\cite{Singh:2020rht} lies in the presence of finite spacetime support and a causal edge, which qualitatively modifies how spin component localizes and mixes during the expansion.

Our results establish the $\kappa=-1$ flow as a qualitatively distinct benchmark for studying spin dynamics in relativistic fluids. Because of its finite spacetime support, this background may be particularly useful for exploring boundary effects, freeze-out prescriptions~\cite{Palermo:2025imv} and the role of strong gradients near the edge of an expanding droplet. Future work could extend the present analysis by including dissipative corrections~\cite{Bhadury:2020cop,Weickgenannt:2022qvh,Daher:2025pfq}, back-reaction of the spin sector on the hydrodynamic background~\cite{Das:2022azr,Drogosz:2024lkx}, the inclusion of electromagnetic fields~\cite{Singh:2021man,Bhadury:2022ulr,Singh:2022ltu,Buzzegoli:2022qrr,Peng:2022cya,Kiamari:2023fbe,Sahoo:2024yud,Fang:2024sym} or coupling to hadronic observables~\cite{Becattini:2021iol,Fu:2021pok,Florkowski:2021wvk,Ivanov:2025izv}. Another theoretical investigation could be to find the mapping between our results for $\kappa=-1$ and Carroll hydrodynamics with spin, which has recently been formulated~\cite{Shukla:2026xig}.

The study of relativistic spin hydrodynamics on maximally symmetric expanding backgrounds has implications that extend well beyond the specific analytic solutions presented here. It sheds light on how spacetime geometry, causal structure, and global symmetries shape the transport and survival of spin polarization in relativistic many-body systems. These questions are of central interest across high-energy nuclear physics, relativistic kinetic theory, and emerging applications of relativistic hydrodynamics in condensed matter systems. We would like to conclude with a broader outlook on the main physical implications of the current work in the following directions:

 \ding{104} \emph{Geometry and causality as organizing principles for spin transport}: A key conceptual implication of this work is that spin dynamics is highly sensitive to the global geometric and causal properties of the underlying flow, not merely to local vorticity or acceleration. In relativistic fluids, spin polarization is often interpreted as a local response to thermal vorticity or shear. However, analytic solutions on curved or conformally related backgrounds demonstrate that the global spacetime structure can qualitatively alter how spin degrees of freedom evolve, mix, and localize.

In particular, expanding geometries with finite spacetime extent provide a controlled setting to disentangle local spin-vorticity coupling from genuinely nonlocal effects induced by expansion rate, curvature, and causal boundaries. This highlights the importance of treating spin polarization as a dynamical field rather than a purely kinematic byproduct of fluid motion.

 \ding{104} \emph{Implications for heavy-ion phenomenology}: In the context of ultra-relativistic heavy-ion collisions, these results reinforce the view that spin polarization observables retain sensitivity to early-time dynamics and global geometry, even when the system subsequently undergoes strong collective expansion. Analytic studies of spin hydrodynamics on nontrivial expanding backgrounds clarify how rapid early-time dilution, transient component mixing and finite-size effects may shape polarization patterns before freeze-out. From a phenomenological standpoint, this suggests that spin observables can encode information about the global spacetime evolution of the medium, complementing traditional flow observables that are primarily sensitive to late-time collective behavior.
More generally, the ability to classify spin dynamics according to symmetry classes of hydrodynamic flows opens the door to systematic benchmarking of spin polarization mechanisms, independent of detailed numerical modeling.

\ding{104} \emph{Constraints on effective theories of spin hydrodynamics}: Analytic solutions on curved backgrounds provide stringent consistency checks for formulations of relativistic spin hydrodynamics. Since spin transport equations couple to geometry through covariant derivatives and connection terms, they are particularly sensitive to assumptions about pseudogauge choice, constitutive relations and the treatment of angular momentum conservation.

The broader implication is that not all formulations of spin hydrodynamics are equally compatible with highly symmetric expanding backgrounds, even at the level of perfect fluids. This motivates further scrutiny of the foundational structure of spin hydrodynamic theories and highlights the value of analytic solutions as diagnostic tools for assessing their internal consistency.

 \ding{104} \emph{Relevance beyond high-energy nuclear physics}: The implications of this work are not confined to heavy-ion collisions. Relativistic hydrodynamics (with spin) has recently attracted attention in condensed matter systems such as Dirac and Weyl materials, where quasiparticles exhibit relativistic dispersion and spin-orbit coupling plays a central role~\cite{Fritz:2008zhk,Mueller:2009guw,Narozhny_2017,Lucas:2017idv,Erdmenger:2018svl,Jaiswal:2024urq,Majumdar2025Nature,Nayak:2025rfg}. In these systems, externally driven expansion, strain-induced curvature, or effective horizons can generate conditions analogous to curved spacetime backgrounds.
The present framework therefore contributes to a growing body of work suggesting that spin transport in relativistic fluids is deeply intertwined with geometry, with potential relevance for spintronics, electron hydrodynamics, and non-equilibrium quantum matter.
\begin{acknowledgments}
RS is supported by a postdoctoral fellowship from the West University of Timișoara, Romania, and acknowledges the kind hospitality of the Faculty of Mathematics and Physics, University of Ljubljana, where the foundation of this research was laid. RS also acknowledges enlightening discussions with Sören Schlichting at the GGI school ``Frontiers in Nuclear and Hadronic Physics 2026'' and Matteo Buzzegoli for clarifying comments. 
AS is supported by funding from 
the project N1-0245 of Slovenian Research Agency (ARIS) and through the UL Startup project (UNLOCK) under contract no. SN-ZRD/22-27/510.
\end{acknowledgments}

\appendix
\section{Equations of motion for \texorpdfstring{$S_{\rm ph}^{\a,\b,\g}$}{}}
\label{app:pheno}
Substituting \cref{eq:S} by \cref{eq:Spheno} together with the decompositions \cref{eq:k_decom} and \cref{eq:o_decom} into \cref{eq:Scon}, and making use of the solutions \cref{eq:GTSol,eq:GMuSol}, we then project the resulting tensor equation onto the basis components $\hat{u}_\b \hat{R}_\g$, $\hat{u}_\b \hat{\Phi}_\g$, $\hat{u}_\b \hat{Z}_\g$,  $\hat{\Phi}_\b \hat{Z}_\g$, $\hat{R}_\b \hat{Z}_\g$ and $\hat{R}_\b \hat{\Phi}_\g$. Projecting \cref{eq:Scon} onto the orthonormal basis $\{\hat{u}^\mu, \hat{R}^\mu, \hat{\Phi}^\mu, \hat{Z}^\mu\}$, we obtain six independent evolution equations for the independent components of the spin potential
\beq
\hat{\cC}\,\dot{\hat{a}}_i + \hat{a}_i \left[\dot{\hat{\cC}} + 2\,  \hat{\cC} \coth{\rho}\right] &=& 0 \quad \text{where} \quad i = R, \Phi, Z\,,\\
\hat{\cC}\,\dot{\hat{b}}_j + \hat{b}_j \left[\dot{\hat{\cC}} +4\,\hat{\cC} \coth{\rho}\right] &=& 0 \quad \text{where} \quad j = R, \Phi, Z\,,
\eeq
with analytic solutions
\beq
\hat{a}_i = \hat{a}_{i}^0\, \sinh^{-2} \rho\,, \quad
\hat{b}_j = \hat{b}_{j}^0\,\sinh^{-4} \rho\,,
\eeq
where $\hat{a}_{i}^0$ and $\hat{b}_{j}^0$ are constants of integration.
\bibliography{pv_ref}{}
\bibliographystyle{utphys}
\end{document}